\documentclass [12 pt] {article}

\def\dddot#1{\mathinner{\buildrel\vbox{\kern5pt\hbox{...}}\over{#1}}}
\def\ddddot#1{\mathinner{\buildrel\vbox{\kern5pt\hbox{....}}\over{#1}}}

\catcode`\@=11

\@addtoreset{equation}{section}
\catcode`\@=10

\begin {document}

\begin {center}
{\Large \bf Symmetry analysis and algebraic structures of the Hu-Paz-Zhang Master Equation}\\[3 mm]
{\large RM Morris${}^{\dagger}$ \& PGL Leach${}^{\dagger \ddagger}$ }\\[3 mm]
{\large {$\dagger$ Department of Mathematics and Institute of Systems Science, Research and Postgraduate Support, Durban University of Technology, PO Box 1334, Durban 4000, Republic of South Africa}\\
{$\ddagger$ School of Mathematics, Statistics and Computer Science, University of KwaZulu-Natal, Private Bag X54001,\\ Durban 4000, Republic of South Africa, and\\
Department of Mathematics and Statistics,\\ University of Cyprus, Lefkosia 1678, Cyprus}}\\[2 mm ]
Email:  \makeatletter rmcalc85@gmail.com; leach@ucy.ac.cy \makeatother
\end {center}
 
\begin{abstract}\noindent
We apply the Lie Theory of continuous groups to investigate the symmetries of the Hu-Paz-Zhang Master Equation which arises in the modelling of the interaction of a harmonic oscillator with a linear passive heat bath of oscillators. We examine the case in which the parameters of the equation are constant.
\end{abstract}

{\bf PACS 2010:} 02.20.Sv; 02.30.Jr; 02.70.Wz

\section {Introduction}

Hu, Paz and Zhang derived an $(1+2)$ evolution equation with the use of path integral methods \cite{hu1,hu2} that inherits the properties of quantum Brownian motion  for a harmonic oscillator interacting with a linear passive heat bath of oscillators. This equation is widely known as the Hu-Paz-Zhang Master Equation. Halliwell and Yu illustrated an alternate and elementary derivation of this equation by tracing the evolution equation for the Wigner Function \cite{halli}. Ford and O'Connell constructed their results on the basis of the quantum Langevin Equation \cite{ford1,ford2}. This equation is given by
\begin{equation}\label {1.1}
Z_{t}=-\frac{x}{m}Z_{y}+m \Omega^{2}(t)y Z_{x}+2 \Gamma (t) (xZ)_{x}+am \Gamma (t) h(t) Z_{xx}+a \Gamma (t) f(t) Z_{xy}, 
\end {equation}
where $Z(t,x,y)$, $a$ and $m$ are real constants and the time-dependent coefficients are given by explicit expressions \cite{ford2,halli}.
\\[2mm]
The integration of (\ref{1.1}) is a nontrivial and cumbersome task. Previous methods that were implemented to determine a solution to (\ref{1.1}) are either numerical or use a Green's Function.  Neither approach is completely satisfactory. The study and analysis of differential equations through the realm of group theory is associated with the great mathematician Sophus Lie \cite{lie} and is a standard but quite effective technique. The advantage of the Lie theory of continuous groups \cite{bluman,ibrag,Olver} is that it allows for greater generality and, even if eventually recourse must be made to numerical methods, this is at the final stage of the calculation and not {\it ab initio}. Such analysis was illustrated in \cite{leach}.

\section {Symmetry analysis}

We analyse the equation given by 
\begin {equation}
Ru(t,x,y)-xu_{y}+Rxu_{x}+ySu_{x}+Vu_{xy}+Wu_{xx}-u_{t}=0 \label {2.1}
\end {equation}
in which there has been a replacement of time-dependent parameters with constant parameters and some renaming and rescaling. 

We determine the Lie point symmetries admitted by (\ref{2.1}) using the add-on package SYM \cite{Andriopoulos09a,Dimas05a,Dimas06a,Dimas08a} associated with the symbolic manipulator Mathematica. 

We take the Lie point symmetry to be of the form
\begin{equation}\label{2.2}
\Delta = \xi^{1}(t,x,y,u)\partial_{t}+\xi^{2}(t,x,y,u)\partial_{x}+\xi^{3}(t,x,y,u)\partial_{y}+\eta(t,x,y,u)\partial_{u}
\end{equation}
and solve the equation
\begin{equation}\label{2.3}
\Delta^{[2]}\left(Ru(t,x,y)-xu_{y}+Rxu_{x}+ySu_{x}+Vu_{xy}+Wu_{xx}-u_{t}\right)=0,
\end{equation}
where $\Delta^{[2]}$ is the standard second extension of $\Delta$.

The analysis of the determining equations is somewhat complicated and we simply summarise the results. As (\ref{2.1}) is linear, it has an infinite number of solution symmetries which we do not take into consideration. 

The symmetries are 
\begin{eqnarray}\label{2.4}
\Delta_{1} &=& \partial_{t}+u\partial_{u},
\nonumber\\[1\jot]
\Delta_{2} &=& u\partial_{u},
\nonumber\\[1\jot]
\Delta_{3} &=& A_{1}(t)\partial_{x}+A_{2}(t)\partial_{y},
\nonumber\\[1\jot]
\Delta_{4} &=& B_{1}(t)\partial_{x}+B_{2}(t)\partial_{y},
\nonumber\\[1\jot]
\Delta_{5} &=& C(t)\left(C_{1}\partial_{x}+C_{2}\partial_{y}+\left(C_{3}x+C_{4}y\right)u\partial_{u}\right),
\nonumber\\[1\jot]
\Delta_{6} &=& E(t)\left(E_{1}\partial_{x}+E_{2}\partial_{y}+\left(E_{3}x+E_{4}y\right)u\partial_{u}\right),
\end{eqnarray}
where $A_{1}(t)$, $A_{2}(t)$, $B_{1}(t)$, $B_{2}(t)$, $C(t)$, ($C_{1}$-$C_{4}$), $E(t)$, ($E_{1}$-$E_{4}$) are given by
\begin{eqnarray*} 
A_{1}(t) &=& -\frac{1}{2}\left(R+\sqrt{R^2-4 S}\right) \exp \left[-\frac{1}{2}\left(R+\sqrt{R^2-4 S}\right)t\right],
\nonumber\\[1\jot]
A_{2}(t) &=& \exp \left[-\frac{1}{2}\left(R+\sqrt{R^2-4 S}\right)t\right],
\end{eqnarray*}

\begin{eqnarray*}
B_{1}(t) &=& \frac{1}{2}\left(-R+\sqrt{R^2-4 S}\right) \exp \left[\frac{1}{2}\left(-R+\sqrt{R^2-4 S}\right)t\right],
\nonumber\\[1\jot]
B_{2}(t) &=& \exp \left[\frac{1}{2}\left(-R+\sqrt{R^2-4 S}\right)t\right],
\end{eqnarray*}

\begin{eqnarray*}
C(t) &=& \frac{1}{2 (RV+W)} \exp \left[-\frac{1}{2}\left(-R+\sqrt{R^2-4 S}\right)t\right],
\nonumber\\[1\jot]
C_{1} &=& R-\sqrt{R^2-4 S}W,
\nonumber\\[1\jot]
C_{2} &=& 2\left(RV+W\right),
\nonumber\\[1\jot]
C_{3} &=& R\left(-R+\sqrt{R^2-4 S}\right),
\nonumber\\[1\jot]
C_{4} &=& -2RS
\end{eqnarray*}
and
\begin{eqnarray*}
E(t) &=& \frac{1}{2 (RV+W)} \exp \left[\frac{1}{2}\left(R+\sqrt{R^2-4 S}\right)t\right],
\nonumber\\[1\jot]
E_{1} &=& R+\sqrt{R^2-4 S}W,
\nonumber\\[1\jot]
E_{2} &=& 2\left(RV+W\right),
\nonumber\\[1\jot]
E_{3} &=& R\left(-R-\sqrt{R^2-4 S}\right),
\nonumber\\[1\jot]
E_{4} &=& -2RS.
\end{eqnarray*}

$\Delta_{1}$ represents the autonomy of (\ref{2.1}) and $\Delta_{2}$ its homogenuity. Neither are very suitable for reduction of order. The structures of ($\Delta_{3}$-$\Delta_{6}$) indicate that they are what has been termed the solution symmetries due to their close connection to the solution symmetries of the corresponding Euler-Lagrange Equation \cite{Andriopoulos05a}. In addition we have the relation $R^2-4S\geq 0$.

In a normal course of events we would replace $\Delta_{1}$ by simply $\partial_{t}$. However, it is instructive to look at the reduction using $2\partial_{t}+Ru\partial_{u}$. The transformation is
\begin{equation}\label{2.5}
u(t,x,y)=\exp \left[\frac{Rt}{2}z(x,y)\right],
\end{equation}
where $z$ satisfies the second-order equation
\begin{equation}\label{2.6}
\frac{1}{2}Rz-xz_{y}+Rxz_{x}+Syz_{x}+Vz_{xy}+W z_{xx}=0.
\end{equation}

\section{Reduction of order to an (1+1) equation}

$\Delta_{3}$: The characteristics are 
\begin{equation}\label{3.1}
u,t \quad\mbox{and}\quad r = \left(R+\sqrt{R^2-4 S}\right)y+2x
\end{equation}
and the reduced equation is
\begin{equation}\label{3.2}
Rz+\frac{1}{2}\left(R-\sqrt{R^2-4 S}\right)rz_{r}+2\left(V\left(R+\sqrt{R^2-4 S}\right)+2W\right)z_{rr}-z_{t}=0.
\end{equation}

The symmetries of (\ref{3.2}) have the form
\begin{equation}\label{3.3}
\Sigma = a(t)\partial_{t}+\left[b(t)+\frac{1}{2}a'(t)r\right]\partial_{r}+z\left(f(t)+F(r,a,b)\right)\partial_{z},
\end{equation}
excluding the solution symmetries, $\phi(t,r)\partial_{z}$, where $a$, $b$ and $f$ satisfy a third-order, a second-order and a first-order equation respectively. Thus the reduced equation is related to the classical Heat Equation by a point transformation.

$\Delta_{4}$: The characteristics are 
\begin{equation}\label{3.4}
u,t \quad\mbox{and}\quad r = \left(-R+\sqrt{R^2-4 S}\right)y-2x
\end{equation}
and the reduced equation is
\begin{equation}\label{3.5}
Rz+\frac{1}{2}\left(R+\sqrt{R^2-4 S}\right)rz_{r}+2\left(V\left(R-\sqrt{R^2-4 S}\right)+2W\right)z_{rr}-z_{t}=0.
\end{equation}

Consequently this also is an $(1+1)$ equation of maximal symmetry.

$\Delta_{5}$: The characteristics are 
\begin{equation}\label{3.6}
t, r = K_{1}Wy-x \quad\mbox{and}\quad u = z \exp \left[(K_{1}Wy-x)RK_{1}y-\frac{1}{2}R(K_{1}^{2}W+K_{2})y^{2}\right],
\end{equation}
where 
\begin{eqnarray*}
K_{1}=\frac{R-\sqrt{R^2-4 S}}{2\left(RV+W\right)} \quad \mbox{and}\quad K_{2}=\frac{S}{\left(RV+W\right)},
\end{eqnarray*}
and, the reduced equation is
\begin{eqnarray}\label{3.7}
&& R\left(r^{2}\left(R-\sqrt{R^2-4 S}\right)+V\left(R+\sqrt{R^2-4 S}\right)+2W\right)z+r\left(V\left(R^{2} \right. \right.
\nonumber\\[1\jot]
&& \left. \left. +\; R\sqrt{R^2-4 S}\right)+W\left(3R-\sqrt{R^2-4 S}\right)\right)z_{r}+\left(VW\left(R+\sqrt{R^2-4 S}\right) \right.
\nonumber\\[1\jot]
&& \left. +\; 2W^{2}\right)z_{rr}-2\left(RV+W\right)z_{t}=0.
\end{eqnarray}

As in the case of (\ref{3.2}) we find that the reduced equation has a maximal number of Lie point symmetries for an $(1+1)$ evolution equation and so is a variant of the classical Heat Equation.

$\Delta_{6}$: The characteristics are 
\begin{equation}\label{3.8}
t, r = K_{3}Wy-x \quad\mbox{and}\quad u = z \exp \left[(K_{3}Wy-x)RK_{3}y-\frac{1}{2}R(K_{3}^{2}W+K_{2})y^{2}\right],
\end{equation}
where 
\begin{eqnarray*}
K_{3}=\frac{R+\sqrt{R^2-4 S}}{2\left(RV+W\right)} 
\end{eqnarray*}
and $K_{2}$ is given as above.

The reduced equation is
\begin{eqnarray}\label{3.9}
&& R\left(r^{2}\left(R+\sqrt{R^2-4 S}\right)+V\left(R-\sqrt{R^2-4 S}\right)+2W\right)z+r\left(V\left(R^{2} \right. \right.
\nonumber\\[1\jot]
&& \left. \left. -\; R\sqrt{R^2-4 S}\right)+W\left(3R+\sqrt{R^2-4 S}\right)\right)z_{r}+\left(VW\left(R-\sqrt{R^2-4 S}\right) \right.
\nonumber\\[1\jot]
&& \left. +\; 2W^{2}\right)z_{rr}-2\left(RV+W\right)z_{t}=0
\end{eqnarray}
and thus consequently also an $(1+1)$ equation of maximal symmetry.

\section{Discussion}

We have seen that in all four cases reductions with the solution symmetries of (\ref{2.1}) leads to an $(1+1)$ evolution equation of maximal symmetry and so is related to the classical Heat Equation by a point transformation. Interestingly in the reduction from the $(1+2)$ equation to the $(1+1)$ equation the number of Lie point symmetries remains the same even though the algebra is different. The $5+1+\infty$ symmetries of (\ref{2.1}) have the algebraic structure $\{A_{1}\oplus_{s}W_{5}\}\oplus_{s}\infty A_{1}$ in which the infinite-dimensional subalgebra comprises the solution symmetries of the linear $(1+2)$ equation (\ref{2.1}), the one-dimensional abelian subalgebra is $\Delta_{1}$ and $W_{5}$ is the five-dimensional Weyl-Heisenberg algebra.
\\[2mm]
In the case of each of the four solution symmetries, $\Delta_{3}$-$\Delta_{6}$, the reduced $(1+1)$ evolution equation possesses $3+3+\infty$ Lie point symmetries with the algebraic structure $\{sl(2,\Re)\oplus_{s}W_{3}\}\oplus_{s}\infty A_{1}$ in which $W_{3}$ is the three-element Weyl-Heisenberg algebra. Usually one sees $5+1+\infty$ and the $5$ relates to the Noether symmetries of the corresponding classical Lagrangian. Here we have preferred to write the former structure to emphasis that the finite algebra comprises two subalgebras each of three elements.
\\[2mm]
Note that in the above we are using the Mubarakzyanov Classification Scheme \cite{Morozov58a,Mubarakzyanov63a,Mubarakzyanov63b,Mubarakzyanov63c} but with some of the algebras listed in the scheme replaced by their common physical names thus $W_{3}$ is $A_{3,3}$ and $sl(2,\Re)$ is $A_{3,8}$, where the latter are the names in the Mubarakzyanov Classification Scheme.
\\[2mm]
It is interesting to note that the same algebraic structures are to be found in the algebraic analysis \cite{soph} of the one and two-factor models introduced by Schwartz \cite{schwartz} in his modelling of the pricing of commodities. Consequently we have the possibility to transform from the physical situation of a thermal bath of oscillators to the commodities market. 
 
\section{Conclusion}

In this paper we have concentrated upon the autonomous form of the Hu-Paz-Zhang Master Equation. Naturally a complete analysis requires examination of the nonautonomous form and we expect that to be reported shortly. In fact there is a number of problems with explicit dependence upon time which are candidates for investigation. These are equations containing parameters which are assumed to be constant for ease of analysis, but which, if one seeks a closer union with reality, should really be functions of time. A recent example of successful analysis is that of the Black-Scholes Equation with arbitrary time-dependent coefficients. The algebra was unchanged \cite{Tamizhmani14a}. One could speculate that this would be the case for the Hu-Paz-Zhang Master Equation.

\section*{Acknowledgements}

RMM thanks the National Research Foundation of the Republic of South Africa for the granting of a postdoctoral fellowship with grant number 85272 while this work was being undertaken. RMM and PGLL thank the Westbrook Beach Club and Estate Manager, Trevor Crichton, for the provision of facilities whilst most of this work was performed.

\end{document}